\documentclass[twocolumn,showpacs,floatfix,preprintnumbers,prl]{revtex4}
\usepackage{amssymb}
\usepackage{graphicx}
\usepackage{dcolumn}
\usepackage{bm}

\begin{document}

\title{Temperature dependence of the angle resolved photoemission spectra 
in the undoped cuprates: self-consistent approach to the t-J-Holstein model}
\author{
 V.~Cataudella$^1$, G.~De~Filippis$^1$, A.~S.~Mishchenko$^{2,3}$ 
and N.~Nagaosa$^{2,4}$}
\affiliation{
$^1$Coherentia-INFM and Dip. di Scienze Fisiche - Universit\`{a} di Napoli 
Federico II - I-80126 Napoli, Italy
\\
$^2$\mbox{CREST, Japan Science and Technology Agency (JST), 
AIST, 1-1-1, Higashi, Tsukuba 305-8562, Japan} 
\\
$^3$RRC ``Kurchatov Institute", 123182, Moscow, Russia 
\\
$^4$CREST, Department of Applied Physics, The University of Tokyo, 7-
3-1 Hongo, Bunkyo-ku, Tokyo 113, Japan
}

\begin{abstract}
We develop a novel self-consistent approach for studying the angle resolved 
photoemission spectra (ARPES) of a hole in the t-J-Holstein model giving 
perfect agreement with numerically exact Diagrammatic Monte Carlo data 
at zero temperature for all regimes of electron-phonon coupling.   
Generalizing the approach to finite temperatures we find that the anomalous 
temperature dependence of the ARPES in undoped cuprates is explained by 
cooperative interplay of coupling of the hole to magnetic fluctuations 
and strong electron-phonon interaction.  
\end{abstract}

\maketitle

Polaron is one of the fundamental problems extensively studied 
both theoretically and experimentally for a long term \cite{polaron}.
However, most of the theoretical works are still restricted to
some limiting cases, or rely on unjustified approximations.
This situation has been recently improved dramatically 
by the combination of the rapid advances in the angle resolved 
photoemission spectroscopy \cite{Shen_03} 
and the numerically exact 
Diagrammatic Monte Carlo (DMC) simulation combined with the 
stochastic analytic continuation \cite{mish_1}. 

Especially the ARPES in the parent compounds of the high 
temperature superconductors, i.e., undoped cuprates, 
turned out to be an ideal arena for studying the
dynamical properties of polaron formation of the single 
hole in the antiferromagnetic background. This is expected from
the ionic nature of the parent compounds and the associated 
strong electron-phonon interaction (EPI), together with the strong electron
correlation as evidenced by the realized Mott insulating state.
Therefore, the interplay between the magnetism and EPI 
is a key to resolve the quantum dynamics of the doped carrier
into the cuprates.

This problem has been theoretically studied \cite{mish_1,rosh} 
in terms of a hole in the t-J model 
coupled by short range Holstein interaction to optical phonons
(t-J-Holstein model). 
The theoretical predictions were verified experimentally 
\cite{shen_2004,shen_2007} and it was shown that the real quasi-particle peak 
has only a tiny weight at lower binding energy compared with the involving 
multi-phonon excitations Franck-Condon peak. 
The former one can be hardly observed because of tiny spectral weight while 
the latter broad one reproduces the dispersion of the pure t-J model.

It has been shown \cite{mish_1,GuSCBA} that the polaronic effect is enhanced 
by the entanglement of the interactions of a hole with magnons and phonons, 
and this interplay is the unique feature of the cuprates. 
Indeed, the EPI alone is absolutely unable to explain the temperature 
dependence of ARPES because experimentally found temperature dependence 
of ARPES is  considerably larger than that predicted by polaronic theory
\cite{kim}. 
A magnetic subsystem alone is also not a suitable candidate since the 
typical energy scale of magnons $\sim 2J \approx 0.2eV$ is even larger than
that of phonons $\sim \omega_0 \approx 0.04eV$. 
Given such desperate situation, there is a temptation to explain the temperature driven 
peak broadening by the approaching of the system to the Neel temperature where 
quantum/thermal fluctuations destroy the antiferromagnetic background of the t-J 
model. 
Recent studies revealed one more puzzle of the temperature dependence 
questioning the polaronic scenario  \cite{mish_1,rosh,GuSCBA}.
The temperature dependence of the line-width is linear in the range 
400K$<T<$200K \cite{shen_2007} and extrapolates to zero line-width at 
zero temperature. From the theoretical point of view, it is a challenge
to study the temperature dependence of the Lehman spectral function (LSF)
for the t-J-Holstein model 
in the intermediate or strong coupling regime in a reliable way, 
which has never been achieved to the best of our knowledge. 

In the present Letter we solve t-J-Holstein model by a novel Hybrid Dynamical 
Momentum Average (HDMA) self-consistent method uniting the advantages of 
Momentum Average (MA) approach \cite{berciu}, keeping the essential 
information on the magnon dispersion, and Dynamical Mean Field (DMF) 
technique, properly taking into account strong but essentially local 
coupling to the lattice.     
Comparing results of HDMA method with exact data obtained by 
DMC approach we show that HDMA method provides accurate 
results for t-J-Holstein model where quasi-particle weakly interacts with 
delocalized magnons and is strongly coupled to local vibrations. 
Making a generalization of HDMA technique to finite temperatures we show 
that the basic features of anomalous temperature dependence of ARPES in 
undoped cuprates can be explained by mutual interplay of magnetic and 
lattice systems in the t-J-Holstein model.      

The Hamiltonian of the t-J-Holstein model in the spin-wave approximation
\cite{varma,kane,horsh_0} reads
\begin{eqnarray}
&&H=\omega _{0}\sum_{\boldsymbol{k}}b_{\boldsymbol{k}}^{\dagger }b_{%
\boldsymbol{k}}+g\omega _{0}\sum_{\boldsymbol{k},%
\boldsymbol{q}}\left[ h_{\boldsymbol{k}}^{\dagger }h_{\boldsymbol{k-q}}b_{%
\boldsymbol{q}}+H.c.\right]   \nonumber \\
&&+\sum_{\boldsymbol{k}}\omega _{\boldsymbol{k}}a_{\boldsymbol{k}}^{\dagger
}a_{\boldsymbol{k}}+\sum_{\boldsymbol{k},\boldsymbol{q}}%
\left[ M_{\boldsymbol{k},\boldsymbol{q}} h_{\boldsymbol{k}}^{\dagger }h_{%
\boldsymbol{k-q}}a_{\boldsymbol{q}}+H.c.\right]   \label{H}
\end{eqnarray}%
where $h_{\boldsymbol{k}}^{\dagger },$ $a_{\boldsymbol{k}}^{\dagger }$ and 
$b_{\boldsymbol{k}}^{\dagger }$ are the creation operators of a hole, a magnon 
and a phonon of momentum $\boldsymbol{k}$, respectively. 
The hole motion is associated with the creation and annihilation of magnons 
of energy $\omega _{%
\boldsymbol{k}}=2J\sqrt{1-\gamma _{\boldsymbol{k}}^{2}}$ ($\gamma _{%
\boldsymbol{k}}=(\cos k_{x}+\cos k_{y})/2$) with coupling 
vertex $M_{\boldsymbol{k},\boldsymbol{q}%
}=4t(u_{q}\gamma _{\boldsymbol{k-q}}+v_{q}\gamma _{\boldsymbol{k}})/\sqrt{N}$
where $%
u_{\boldsymbol{q}}=\sqrt{(1+\alpha _{q})/(2\alpha _{q})}$, $v_{\boldsymbol{q}%
}=-sgn(\gamma _{\boldsymbol{q}})\sqrt{(1-\alpha _{q})/(2\alpha _{\boldsymbol{%
q}})}$, $\alpha _{\boldsymbol{q}}=\omega _{\boldsymbol{q}}/2J$, and $N$ is
the number of lattice sites. 
The short-range interaction between the hole and local distortions
due to dispersionless optical vibrations with frequency $\omega _{0}$ 
is described by the coupling constant $g$. For the following we use 
the corresponding to experiment values $J/t=0.3$, $\omega_0/t=0.1$ 
\cite{Shen_03}, measure all energies in units of $t$ and assume Planck and 
Boltzmann constants equal to unity.

The generic features of the model (\ref{H}), causing difficulties to 
semi-analytic approaches, is the intrinsic interplay between interaction
of a hole with magnons reducing the spectral weight of its quasi-particle as
well as reducing its bandwidth and coupling to local phonons backing the
self-trapping of the quasi-particle.  
Brutal force disentangling of these two contributions is impossible
because the energy scales of two processes are of the same order 
\cite{horsh_2006} and the only attempts which were successful so far, in 
quantitative description of the spectral properties of the model (\ref{H})  
at zero temperature, were based on numerically involved methods, 
such as exact diagonalization \cite{horsh_2006,fehske_1} or DMC \cite{mish_1}
techniques. 
However, the results of the former method are limited to small lattices
while the latter one, working in the thermodynamic limit, requires extremely 
extensive numerics efforts at finite temperature due to the ``ill posed'' 
nature of the analytic continuation \cite{MPSS}.  

In spite of the same energy scales involved into the hole-magnon and 
hole-phonon couplings, these two interactions are profoundly different 
since the coupling to magnons is essentially momentum dependent and 
always weak whereas that to phonons is local and can be strong. 
Indeed,  spin $S=1/2$ cannot flip more than one time assuring that each site 
can not possess more than one magnon \cite{barentzen} and, thus, 
the weak-coupling Self-Consistent Born Approximation (SCBA) is satisfactory
for small values of $J/t$ \cite{manusakis,GuSCBA}.  
To the contrary, SCBA fails for EPI even in the intermediate coupling limit 
\cite{mish_1}. 
Therefore, to cope with t-J-Holstein problem it is enough to treat the 
essential momentum dependent coupling to magnons within the SCBA and to sum 
vibrational variables nonperturbatively, at least in some local approximation.
Nonperturbative local approaches, valid at any coupling strength and 
neglecting the $\boldsymbol{k}$-dependence in the self-energy 
$\Sigma_{h\_ph}(\boldsymbol{k},\omega )$, are 
DMF technique \cite{dmf} and recently developed MA 
method \cite{berciu} providing explicit form for the hole self-energy due to hole-phonon interaction  
in terms of a continued fraction 
\begin{equation}
\Sigma _{\text{h-ph}}[\alpha (\omega )]=\frac{(g\omega _{0})^{2}\alpha
(\omega -\omega _{0})}{1-\frac{2(g\omega _{0})^{2}\alpha (\omega -\omega
_{0})\alpha (\omega -2\omega _{0})}{1-\frac{3(g\omega _{0})^{2}\alpha
(\omega -2\omega _{0})\alpha (\omega -3\omega _{0})}{1-...}}} \; .
\label{sigma_ph}
\end{equation}%
The difference in DMF and MA lies in the definition of $\alpha (\omega )$ 
which is a function that has to be fixed by a self-consistent procedure in the 
DMF approach while is identified with the $\boldsymbol{k}$-average of the 
bare Green's function in the MA scheme. 
Obviously, MA scheme is preferable when one is interested in properties 
of 2D model (\ref{H}) with highly anisotropic coupling  
 $M_{\boldsymbol{k},\boldsymbol{q}}$. 

Summarizing the above considerations a reasonable self-consistent 
procedure expresses the total self-energy of the hole as the the sum of the 
self-energies caused by magnetic and phonon subsystems
\begin{equation}
\Sigma _{tJH}(\boldsymbol{k},\omega )=\Sigma _{\text{h-mag}}^{SCBA}(%
\boldsymbol{k},\omega )+\Sigma _{\text{h-ph}}[\alpha _{tJH}(\omega )].
\label{sigma}
\end{equation}%
Weak and highly anisotropic interaction with magnons is taken into 
account in the SCBA 
\begin{equation}
\Sigma _{\text{h-mag}}^{SCBA}(\boldsymbol{k},\omega )=  \nonumber
\sum_{\boldsymbol{q}}\frac{M_{\boldsymbol{k},\boldsymbol{q}}^{2}%
}{\omega -\omega _{\mathbf{q}}-\Sigma _{tJH}(\boldsymbol{k-q},\omega -\omega
_{\mathbf{q}})+i\varepsilon }  \label{sigma_mag}
\end{equation}
and the $\alpha(\omega)$-function for hole-phonon self-energy  
\begin{equation}
\alpha _{tJH}(\omega )=\frac{1}{N}\sum_{\boldsymbol{k}}\frac{1}{\omega
-\Sigma _{\text{h-mag}}^{SCBA}(\boldsymbol{k},\omega )+i\varepsilon }
\label{alpha}
\end{equation}
is expressed in terms of momentum average of ``bare'' Green function whose
${\bf k}$-dependence is determined by the hole-magnon self-energy 
(\ref{sigma_mag}) in the SCBA. 
The equations (\ref{sigma_ph}-\ref{alpha}) constitute the self-consistent set
of the HDMA approach and can be solved by standard iterative methods on a 
finite lattice by evaluating $\omega $ on a finite mesh of points.

\begin{figure}[tbp]
\includegraphics[scale=0.9]{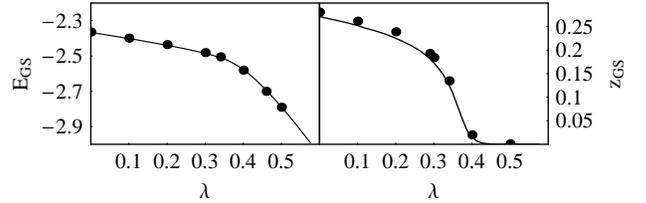}
\caption{The ground state energy (E$_{GS}$) and the spectral
weight ($z_{GS}$) as a function of the EPI strength 
$\lambda=g^{2}\omega _{0}/4t$
in the DMC approach (line) and in the HDMA method (points).}
\end{figure}

The set of eqs.\ (\ref{sigma_ph}-\ref{alpha}) has the same structure 
as obtained in the DMF formulation of the t-J-Holstein model 
\cite{ciuchi_tJH}, with the important exception that the 
$\alpha(\omega)$-function is determined not from the purely local 
self-consistent DMF condition but defined through the momentum average 
\cite{berciu} of the ``bare'' Green function containing the anisotropic 
self-energy of two dimensional t-J model.  
Within the framework of DMF approach the t-J-Holstein model is 
indistinguishable from $t$-$J_{z}$ model where the hole coherent motion is 
suppressed\cite{ciuchi_tJH}. 
To the contrary, HDMA approach preserves coherent motion of the hole.  
The ground state energy, $E_{GS}$, and its spectral weight, 
z$_{GS}=\left( 1-\left. \partial \Sigma _{tJH}/\partial
\omega \right\vert _{\omega =E_{GS}}\right) ^{-1}$ are in good  
agreement with the data of numerically exact DMC approach \cite{mish_1} 
and the crossover to the strong coupling limit at $\lambda \ge 0.4$ is 
perfectly reproduced (see Fig.~1).

\begin{figure}[tbp]
\includegraphics[scale=0.9]{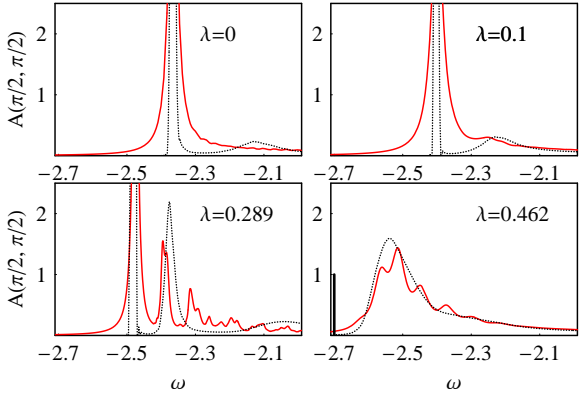}
\caption{The LSF $A(\protect\pi /2,\protect\pi /2)$ for different EPI  
$\lambda$ in the HDMA (solid line) and DMC (dotted line) approaches.
Vertical line in the panel with $\protect\lambda =0.462$ indicates the 
position of the ground state quasi-particle peak.}
\end{figure}

For further check of validity of the HDMA scheme,  we compare the 
spectral function calculated on a lattice $64\times 64$ with that obtained 
by the approximation-free DMC technique for pure t-J model ($\lambda =0$) 
and t-J-Holstein model in the weak ($\lambda =0.1$), intermediate 
($\lambda =0.289$), and strong ($\lambda =0.462$) coupling regimes (Fig.~2). 
The very good agreement of the overall shapes is observed for all 
coupling regimes and mismatch of fine details can be attributed 
to finite size effects of $64\times 64$ lattice 
\footnote{The acoustic spectrum of the magnons is very sensitive to the size 
effects. For instance, at small and intermediate coupling the second peak 
is entirely due to EPI in our approach while DMC data suggest a 
more complex nature where both magnons and phonons contribute.} 
and the "local" approximation used in the present approach 
[eq. (\ref{sigma_ph})] that, at strong coupling, gives the typical 
oscillations with the period of phonon energy.
Apart from these details our approach is reliable in all coupling regimes and 
gives the spectral function with a computational effort much less than spent 
by DMC.

An important advantage of our scheme is that its generalization to 
finite temperature is straightforward. 
Performing analytical continuation of 
$\Sigma _{\text{h-mag}}^{SCBA}(\boldsymbol{k},\omega )$ 
to Matsubara formalism one gets \cite{mahan, sawatzky} 
\begin{eqnarray}
&&\Sigma _{\text{h-mag}}^{SCBA}(\boldsymbol{k}\omega )=  \nonumber
\sum_{\boldsymbol{q}}\frac{M_{\boldsymbol{k},\boldsymbol{q}%
}^{2}(1+n_{b}(\omega _{\mathbf{q}}))}{\omega -\omega _{\mathbf{q}}-\Sigma
_{tJH}(\boldsymbol{k-q},\omega -\omega _{\boldsymbol{q}})+i\varepsilon }+
\nonumber \\
&&\sum_{\boldsymbol{q}}\frac{M_{\boldsymbol{k+q},\boldsymbol{q}%
}^{2}(n_{b}(\omega _{\mathbf{q}}))}{\omega +\omega _{\mathbf{q}}-\Sigma
_{tJH}(\boldsymbol{k+q},\omega +\omega _{\boldsymbol{q}})+i\varepsilon } \; ,
\label{sigma_mag_T}
\end{eqnarray}%
where $n_{b}(\omega )$ is the Bose-Einstein factor. 
For the generalization of Eq.~(\ref{sigma_ph}) one notes that  hole self-energy due to hole-phonon interaction $\Sigma _{h-ph}(\omega )$ for the model
where the hole can interact with phonon only when it is on the site $i$ the 
temperature dependence can be included in the exact way giving 
\cite{cini,ciuchi_0}
\begin{equation}
\Sigma _{\text{h-ph}}[\alpha (\omega )]=\alpha ^{-1}(\omega
)-\sum_{n=0}^{\infty }\frac{(1-x)x^{n}}{\alpha ^{-1}(\omega )-A_{n}(\omega
)-B_{n}(\omega )}  \label{sigma_ph_T}
\label{tempe}
\end{equation}%
where $x=\exp (-\beta \omega _{0})$ and 
\begin{eqnarray*}
A_{n}(\omega ) &=&\frac{n(g\omega _{0})^{2}\alpha (\omega +\omega _{0})}{1-%
\frac{(n-1)(g\omega _{0})^{2}\alpha (\omega +\omega _{0})\alpha (\omega
+2\omega _{0})}{1-\frac{(n-2)(g\omega _{0})^{2}\alpha (\omega +2\omega
_{0})\alpha (\omega +3\omega _{0})}{1-...}}} \\
B_{n}(\omega ) &=&\frac{(n+1)(g\omega _{0})^{2}\alpha (\omega -\omega _{0})}{%
1-\frac{(n+2)(g\omega _{0})^{2}\alpha (\omega -\omega _{0})\alpha (\omega
-2\omega _{0})}{1-\frac{(n+3)(g\omega _{0})^{2}\alpha (\omega -2\omega
_{0})\alpha (\omega -3\omega _{0})}{1-...}}} \; .
\end{eqnarray*}
Since the $\alpha(\omega)$-function in Eq.~(\ref{alpha}) is reduced to a local 
momentum independent value, above expression is also valid for our scheme. 

\begin{figure}[bth]
\includegraphics[scale=0.9]{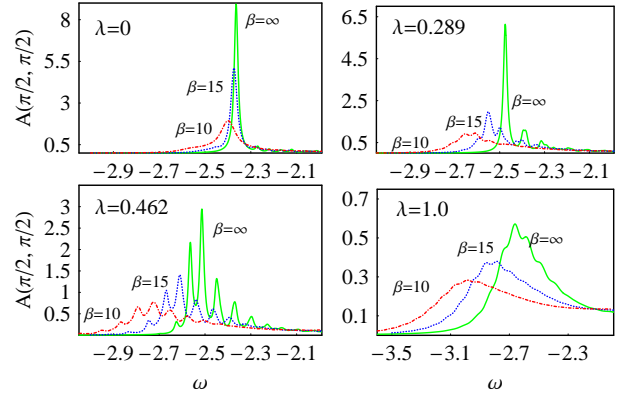}
\caption{The LSF ($A(\protect\pi /2,\protect\pi /2)$) 
for different 
EPI couplings $\lambda$ and different temperatures 
$\protect\beta =t/T$. }
\end{figure}

The Eqs.\ (\ref{sigma},\ref{alpha},\ref{sigma_mag_T},\ref{sigma_ph_T}) 
provide a set of self-consistent equations that can be solved iteratively
typically within 40 iterations.
We verified the relevance of the relation (\ref{tempe}) for our scheme 
checking the sum rules \cite{berciu} and found that the first three sum
rules for the LSF are satisfied at any temperature and EPI with
high accuracy. 
Therefore, our results for peak energy and linewidth, determined mainly by 
the first and second sum rule, do not lose accuracy from the approximations 
made to obtain the HDMA self-consistent scheme.

\begin{figure}[tbp]
\includegraphics[scale=0.8]{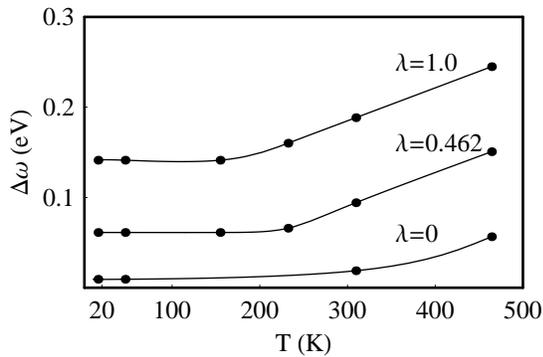}
\caption{The peak's width, $\Delta \protect\omega ,$ as a function of the
temperature, $T,$ for the t-J model ($\protect\lambda =0$) and in 
strong EPI limit ($\protect\lambda =0.462$ and $\lambda =1$). 
The temperature $T$ is defined in units of Kelvin assuming $t=0.4$eV.}
\end{figure}

The temperature dependence of LSF at $k=(\pi /2,\pi /2)$ at different 
values of EPI is shown in Fig.~3. The trends for peak position and 
linewidth are in agreement with experimental data 
\cite{shen_2007,kim,pothui_1,pothui_2}.
With the increase of temperature the binding energy and width of the main 
broad peak increase while its intensity decreases. 
It is seen that all temperature driven effects are more pronounced
in the t-J-Holstein model than in the t-J model supporting the 
statement \cite{mish_1,GuSCBA} that the entanglement of the magnetic and 
vibrational fluctuations is essential for cuprates and crucial for 
description of anomalously enhanced temperature driven effects in 
undoped compounds \cite{kim}.  

The temperature dependence of the peak width estimated through a Gaussian 
fitting is shown in Fig.~4.   
It is remarkable that in the strong coupling regime of the EPI the peak 
width is almost constant up to $T \simeq \omega _{0}/2 \approx 200$K and 
then, for $T\geqq \omega_{0} /2$, demonstrates linear dependence which, 
in according with experiment, can be naively extrapolated to almost zero 
value at zero temperature.  
Note, the temperature dependence is strongly enhanced in the strong coupling 
limit $\lambda>0.4$ of the t-J-Holstein model giving in this limit, in contrast
with purely polaronic or purely magnetic models, correct order of magnitude
of the effect and even showing a good semi-quantitative agreement with 
experiment \cite{shen_2007,kim,pothui_1,pothui_2}. 
For $\lambda=0.462$ the peak width, in quantitative agreement with experiment
\cite{shen_2007}, doubles in the range from 200K to 400K though the absolute
value of the peak width is a factor of 2 below the experimental values. 
On the other hand, the absolute value of the linewidth fits the experiment 
for $\lambda=1$ but the enhancement of the width is a factor of 1.5 below 
that found experimentally \cite{shen_2007}. 
The above discrepancies can be attributed to the fact that the longer range 
hoppings of more realistic tt$'$t$''$-J model are missing in the t-J model 
or as well to the fact that in the realistic systems the holes are 
coupled to several phonon modes through the EPI of different strength
\cite{X-J05}.        

In conclusion, by using a new hybrid dynamic momentum average approach to 
the calculation of a hole LSF in the t-J Holstein model we have shown 
that the origin of the anomalously large  temperature dependence of the ARPES
in the undoped parent compound of high temperature superconductors  
originates from the constructive interplay between magnetic and 
strong electron-phonon interactions.  

We acknowledge fruitful discussions with Prof. Z.X.Shen and Dr. K.M. Shen.
One of the authors (N.N.) acknowledges the financial support from the Grant-in-Aids under the Grant numbers 15104006, 16076205, and 17105002, and NAREGI Nanoscience Project from the Ministry of Education, Culture, Sports, Science, and Technology, Japan. ASM acknowledges support of RFBR 07-0200067-a.

\end{document}